\documentclass[12pt]{article}
\usepackage{amscd,amssymb,epsf}

\advance \topmargin by -\headheight
\advance \topmargin by -\headsep

\evensidemargin \oddsidemargin
\def\ie{{\it i.e.}}

\def\CC{\mathcal{C}}
\def\CP{\mathcal{P}}

\def\Too{T^{1,1}}
\def\Ypq{Y^{p,q}}
\def\CYpq{\mathcal{C}Y^{p,q}}
\def\Labc{L^{a,b,c}}
\def\CLabc{\mathcal{C}L^{a,b,c}}

\newcommand{\beq}{\begin{equation}}
\newcommand{\eeq}{\end{equation}}
\newcommand{\rc}{\nonumber\\}
\newcommand{\bear}{\begin{eqnarray}}
\newcommand{\eear}{\end{eqnarray}}

\begin{document}
%

%
%

\begin{center} \Large \bf D--brane probes on $\Labc$ Superconformal Field
Theories
\end{center}

\vskip 0.3truein
\begin{center}
Felipe Canoura${}^{\,*}$
\footnote{canoura@fpaxp1.usc.es},
Jos\'e D. Edelstein${}^{\,*\dagger}$
\footnote{edels@fpaxp1.usc.es}
and
Alfonso V. Ramallo${}^{\,*}$
\footnote{alfonso@fpaxp1.usc.es}

\vspace{0.3in}

${}^{\,*}$Departamento de F\'\i sica de Part\'\i culas, Universidade de
Santiago de Compostela \\ and
Instituto Galego de F\'\i sica de Altas Enerx\'\i as (IGFAE)\\
E-15782 Santiago de Compostela, Spain
\vspace{0.2in}

${}^{\,\dagger}$Centro de Estudios Cient\'\i ficos (CECS) Casilla 1469,
Valdivia, Chile

\end{center}
\vskip.5truein

\begin{center}
\bf ABSTRACT
\end{center}
We study supersymmetric embeddings of D-brane probes of different
dimensionality in the $AdS_5\times \Labc$ background of type
IIB string theory. In the case of D3-branes, we recover the
known three-cycles dual to the dibaryonic operators of the gauge theory
and we also find a new family of supersymmetric embeddings.
Supersymmetric configurations of D5-branes, representing fractional
branes, and of spacetime filling D7-branes (which can be used to add
flavor) are also found. Stable non supersymmetric configurations
corresponding to fat strings and domain walls are found as well.

\vskip1.3truecm
\leftline{US-FT-3/06 \hfill hep-th/0605260}
\leftline{CECS-PHY-06/14 \hfill 26 May 2006}
\medskip
\setcounter{footnote}{0}


\setcounter{equation}{0}
\section{Introduction}
\label{intro}
\medskip

The study of supersymmetric  D--brane probes in a given background is
relevant to extract {\it stringy} information in the framework of
gauge/gravity duality. This is a well-known fact since Witten early
showed, in the case of ${\cal N}=4$ supersymmetric Yang--Mills (SYM)
theory, that the gravity side must contain branes in order to accommodate
the Pfaffian operator --in the $SO(2N)$ case--, as well as the baryon
vertex or domain walls arising in $SU(N)$ gauge theories \cite{Wi}.
Moreover, the introduction of flavor in the gauge theory side forces to
consider an open string sector in the dual theory \cite{KaKa}. As a
consequence, any theory in the universality class of QCD demands a clear
understanding of these features.

In the long path from Maldacena's original setup to more realistic scenarios,
an important framework has been considered in recent years. If the five-sphere 
of the background
is replaced by any Sasaki--Einstein five-dimensional manifold $X^5$, a duality between type IIB string theory on $AdS_5\times X^5$ and a superconformal quiver
gauge theory arises \cite{Gu}. Until two years ago, the only case at hand whose
complete
metric was known was $X^5 = \Too$, that leads to the so-called 
Klebanov--Witten model \cite{KlWi}. More recently, two new families of
infinitely many Sasaki--Einstein
 manifolds were built and their metrics were explicitly constructed.
They are labeled by two positive integers $\Ypq$ \cite{GaMaSpWa} or three
positive integers $\Labc$ \cite{CvLuPaPo,MaSp2}. Indeed, the former can
be seen as a subfamily of the latter. The corresponding superconformal
field theories were constructed almost immediately in, respectively,
\cite{MaSp,BeFrHaMaSp} and \cite{BeKr,FrHaMaSpVeWe,BuFoZa}, by exploiting
the rich mathematical structure of toric geometry. These families exhaust
all possible toric Calabi--Yau cones on a base with topology $S^2 \times
S^3$. Research on AdS/CFT in these superconformal gauge theories has led
to a better understanding of several important issues such as the
appearance of duality cascades, a--maximization, Seiberg duality, etc.

In order to determine the supersymmetric embeddings of D--brane probes we
employ kappa symmetry \cite{swedes}. Our approach is based on the existence
of a matrix $\Gamma_{\kappa}$ which depends on the metric induced on the worldvolume of the probe and characterizes its supersymmetric embeddings.
If $\epsilon$ is a Killing spinor of the background, only those embeddings
such that $\Gamma_{\kappa}\,\epsilon\,=\,\epsilon$ preserve a fraction of
the background supersymmetry \cite{bbs}. This condition gives rise to a
set of first-order BPS differential equations whose solutions determine
the details of the embedding. As well, they solve the equations of motion derived from the DBI action of the probe while saturating a bound for the energy, as it usually happens in the case of worldvolume solitons \cite{GGT}. 

D-brane probes in the Klebanov--Witten model were studied in full detail
in \cite{ArCrRa}. In the case of $\Ypq$ superconformal gauge theories,
the exhaustive research was undertaken more recently in \cite{CaEdPaZaRaVa}.
These articles explore interesting features such as excitations of dibaryons,
the baryon vertex, the presence of domain walls, fat strings, defect conformal
field theories, in the quiver theory side. In this letter, we aim at filling
the gap by giving the main results in the case of $\Labc$ theories.

The content of this article is organized as follows. In Section 
\ref{Labcgeometry} we review those aspects of the $\Labc$ spaces that
we need afterwards. Section \ref{complexcoordinates} deals with the
construction of local complex coordinates and other geometrical aspects
of the Calabi--Yau cone,
$\CLabc$. In Section \ref{Kspinors} we provide the expression for the
Killing spinors on $AdS_5 \times
\Labc$. We briefly describe the basics of the dual superconformal quiver
theories in Section \ref{Quiver}. We consider D3--branes wrapping
supersymmetric 3-cycles dual to dibaryonic operators in Section \ref{D3s}.
Besides matching their quantum numbers, we find general holomorphic
embeddings corresponding to divisors of $\CLabc$. In Section 
\ref{D5s} we consider
D5--branes with the focus on fractional branes, while Section 
\ref{D7s} deals
with spacetime filling configurations of D7--branes that can be used to
introduce flavor. We finally comment on some stable non-supersymmetric
configurations representing fat strings and domain walls in Section
\ref{FinalRemarks}, where we furthermore present our conclusions.

\setcounter{equation}{0}
\section{The $\Labc$ geometry}
\label{Labcgeometry}
\medskip

The Sasaki--Einstein manifold $\Labc$ is a five-dimensional space with
topology $S^2 \times S^3$, whose metric can be written as \cite{CvLuPaPo}:
\beq
ds^2_{\Labc}\,=\,ds_4^2\,+\,(d\tau\,+\,\sigma)^2\,\,,
\eeq
where $ds_4^2$ is a local K\"ahler--Einstein metric, with K\"ahler form
$J_4 = \frac{1}{2} d\sigma$, given by
\bear
&&ds_4^2\,=\,{\rho^2\over 4\Delta_x}\,dx^2\,+\,
{\rho^2\over \Delta_{\theta}}\,d\theta^2\,+\,
{\Delta_x\over \rho^2}\,\Bigg(\,
{\sin^2\theta\over \alpha}\,d\phi\,+\,{\cos^2\theta\over \beta}\,d\psi
\,\Bigg)^2\,\,+\rc\rc
&&\qquad~~~~~~~~+ \, {\Delta_\theta\sin^2\theta\cos^2\theta\over \rho^2}\,\,
\Bigg[\,\bigg(1-{x\over \alpha}\bigg)\,d\phi\,-\,
\bigg(1-{x\over \beta}\bigg)\,d\psi\,\Bigg]^2\,\,,
\eear
the quantities $\Delta_{x}$, $\Delta_{\theta}$, $\rho^2$ and $\sigma$ reading
\bear
&&\Delta_{x}\,=\,x(\alpha-x)(\beta-x)\,-\,\mu\,\,,\rc\rc
&&\Delta_{\theta}\,=\,\alpha \cos^2\theta \,+\,\beta\sin^2\theta\,\,,
\qquad\rho^2\,=\,\Delta_{\theta}\,-\,x\,\,,\rc\rc
&&\sigma\,=\,\bigg(1-{x\over \alpha}\bigg)\,\sin^2\theta\,d\phi\,+\,
\bigg(1-{x\over \beta}\bigg)\,\cos^2\theta \,d\psi\,\,.
\eear
The ranges of the different coordinates are $0\le \theta\le \pi/2$, $x_1
\le x \le x_2$, $0\le \phi, \psi <2\pi$, where $x_1$ and $x_2$ are the
smallest roots of the cubic equation $\Delta_x=0$. A natural tetrad frame
for this space reads
\bear
&&e^1\,=\,{\rho\over \sqrt{\Delta_{\theta}}}\,d\theta\,\,,\qquad
e^2\,=\,{\sqrt{\Delta_{\theta}}\,\,\sin\theta\cos\theta\over \rho}
\Bigg(\,\bigg(1-{x\over \alpha}\bigg)\,d\phi\,-\,
\bigg(1-{x\over \beta}\bigg)\,d\psi\,\Bigg)\,\,,\rc\rc
&&e^3\,=\,{\sqrt{\Delta_x}\over \rho}\,\,
\Bigg(\,{\sin^2\theta\over \alpha}\,d\phi\,+\,{\cos^2\theta\over
\beta}\,d\psi\,\Bigg)\,\,,\rc\rc
&&e^4\,=\,{\rho\over 2 \sqrt{\Delta_x}}\,\,dx\,\,, \qquad\qquad
e^5\,=\,\big(\,d\tau+\sigma)\,\,.
\eear
Notice that, in this frame, $J_4 = e^1 \wedge e^2 + e^3 \wedge e^4$.
Let us now define $a_i$, $b_i$ and $c_i$ ($i=1,2$) as follows:
\beq
a_i\,=\,{\alpha c_i\over x_i-\alpha}\,\,,\qquad
b_i\,=\,{\beta c_i\over x_i-\beta}\,\,,\qquad
c_i\,=\,{(\alpha-x_i)(\beta-x_i)\over 2(\alpha+\beta)\,x_i\,-\,
\alpha\beta\,-\,3x_i^2}\,\,.
\label{aibici}
\eeq
The coordinate $\tau$ happens to be compact and varies between 0 and $\Delta
\tau$,
\beq
\Delta \tau\,=\,{2\pi k |c_1|\over b}\,\,,
\qquad k={\rm gcd }\,(a,b)\,\,.
\eeq
The $a_i$, $b_i$ and $c_i$ constants are related to the integers $a,b,c$ of
$\Labc$ by means of the relations:
\beq
a\,a_1\,+\,b\,a_2\,+\,c\,=\,0\,\,,\qquad
a\,b_1\,+\,b\,b_2\,+\,d\,=\,0\,\,,\qquad
a\,c_1\,+\,b\,c_2\,=\,0\,\,,
\label{abc-relations}
\eeq
where $d=a+b-c$. The constants $\alpha$, $\beta$
and $\mu$ appearing in the metric are related to the roots $x_1$, $x_2$  and
$x_3 $ of $\Delta_x$ as
\beq
\mu\,=\,x_1x_2x_3\,\,,\qquad
\alpha+\beta\,=\,x_1+x_2+x_3\,\,,\qquad
\alpha\beta\,=\,x_1x_2+x_1x_3+x_2x_3\,\,.
\label{xs}
\eeq
Moreover, it follows from (\ref{abc-relations}) that all ratios bewteen
the four quantities $a_1c_2-a_2c_1$, $b_1c_2-b_2c_1$, $c_1$, and $c_2$
must be rational. Actually, one can prove that:
\beq
{a_1c_2-a_2c_1\over c_1}\,=\,{c\over b}\,\,,\qquad
{b_1c_2-b_2c_1\over c_1}\,=\,{d\over b}\,\,,\qquad
{c_1\over c_2}\,=-\,{b\over a}\,\,.
\label{ratios}
\eeq
Any other ratio between $(a,b,c,d)$ can be obtained by combining these
equations. In particular, from (\ref{aibici}),  (\ref{xs}) and
(\ref{ratios}), one can rewrite some of these relations as:
\bear
&&{a\over b}\,=\,{x_1\over x_2}\,\,{x_3-x_1\over x_3-x_2}\,\,,\qquad
{a\over c}\,=\,{(\alpha-x_2)(x_3-x_1)\over \alpha(\beta-x_1)}\,\,,\rc\rc
&&{c\over d}\,=\,{\alpha\over \beta}\,\,
{(\beta-x_1)(\beta-x_2)\over (\alpha-x_1)(\alpha-x_2)}\,=\,
{\alpha\over \beta}\,\,{x_3-\alpha\over x_3-\beta}\,\,.
\label{ratios2}
\eear
The manifold has $U(1) \times U(1) \times U(1)$ isometry. It is, thus, toric.
Its volume can be computed from the metric with the result:
\beq
{\rm Vol}(\Labc)\,=\,{(x_2-x_1)(\alpha+\beta-x_1-x_2)\,|c_1|\over
\alpha\beta b}\,\,\pi^3\,\,.
\eeq
Other geometrical aspects of these spaces can be found in
\cite{CvLuPaPo,MaSp}.

\setcounter{equation}{0}
\section{Complex coordinates on  ${\cal C}\Labc$ }
\label{complexcoordinates}
\medskip

In order to construct a set of local complex coordinates on the Calabi--Yau
cone on $\Labc$, $\CLabc$, let us introduce the following basis of closed
one-forms \footnote{Notice that there are a few sign differences in our conventions as compared to those in \cite{SfZo}.}
\bear
&&\hat\eta_1\,=\,\alpha \,{\cot\theta\over \Delta_{\theta}}\,d\theta\,-\,
{\alpha (\beta-x)\over 2\Delta_x}\,dx\,+\,id\phi\,\,,\rc\rc
&&\hat\eta_2\,=\,-\beta\,{\tan\theta\over \Delta_\theta}\,d\theta\,-\,
{\beta (\alpha-x)\over 2\Delta_x}\,dx\,+\,id\psi\,\,,\rc\rc
&&\hat\eta_3\,=\,{dr\over r}\,+\,id\tau\,+\,
\big(\beta-\alpha)\,{\sin (2\theta)\over 2\Delta_{\theta}}\,d\theta\,+\,
{(\alpha-x)(\beta-x)\over 2\Delta_x}\,dx\,\,.
\eear
From these quantities, it is possible to define a set of $(1,0)$-forms
$\eta_i$ as the following linear combinations:
\beq
\eta_1\,=\,\hat \eta_1-\hat \eta_2\,\,,\qquad
\eta_2\,=\,\hat \eta_1+\hat \eta_2\,\,,\qquad
\eta_3\,=\,3\hat \eta_3\,+\,\hat \eta_1+\hat \eta_2\,\,.
\eeq
One can immediately check that they are integrable, $\eta^i\,=\,{dz^i\over
z^i}$. The explicit form of the complex coordinates $z^i$ is:
\bear
&&z_1\,=\,\tan\theta\,f_1(x)\,e^{i(\phi-\psi)}\,\,,\qquad
z_2\,=\,{\sin(2\theta)\over
f_2(x)\,\Delta_{\theta}}\,e^{i(\phi+\psi)}\,\,,\rc\rc
&&z_3\,=\,r^3\,\sin(2\theta)\,\sqrt{\Delta_{\theta}\Delta_x}\,\,
e^{i(3\tau+\phi+\psi)}\,\,,
\label{zs}
\eear
where
\beq
f_1(x)\,=\,\CP_1(x)^{\alpha\,-\,\beta}\,\,,\qquad
f_2(x)\,=\,\CP_0(x)^{2\,\alpha\,\beta}\,\CP_1(x)^{-(\alpha\,+\,\beta)}\,\,,
\label{f12}
\eeq
and the functions $\CP_q(x)$ are defined as
\beq
\CP_q(x) = \exp\,\left(\,\int\,{x^q\,dx\over 2\,\Delta_x}\,\right)\,=\,
\prod_{i=1}^3\,(x\,-\,x_i)^{{1\over 2} {x_i^q\over \prod_{j\neq i}^3\,(x_i\,-\,x_j)}}
\,\,.
\eeq
In terms of these $(1,0)$-forms, it is now fairly simple to work out the
two-form $\Omega_4$,
\beq
\Omega_4\,=\,3e^{i(\phi+\psi)}\,\sin(2\theta)
\sqrt{\Delta_\theta\Delta_x}\,\,\,\eta_1\wedge\eta_2\,\,,
\eeq
obeying $d\,\Omega_4\,=\,3i\sigma\wedge \Omega_4$. By using these properties
one can verify that the three-form:
\beq
\Omega\,=\,r^2\,e^{3i\tau}\,\Omega_4\wedge \big[\,dr\,+\,ir\,
(d\tau+\sigma)\,\big]\,\,,
\eeq
is closed. Moreover, the explicit expression for $\Omega$ in terms of the
above defined closed and integrable $(1,0)$-forms reads
\beq
\Omega\,=\,r^3\,\sin(2\theta)\,
e^{i(3\tau+\phi+\psi)}\,\sqrt{\Delta_\theta\,\Delta_x}\,\,
\eta_1\wedge\eta_2\wedge\eta_3\,\,,
\label{OmegaR}
\eeq
which shows that $\Omega\wedge \eta_i=0$. In terms of the complex coordinates $z_i$, the form $\Omega$ adopts a simple expression from which it is clear
that it is the holomorphic (3,0) form of the Calabi-Yau cone $\CLabc$,
\beq
\Omega\,=\,{dz_1\wedge dz_2\wedge dz_3\over z_1 z_2}\,\,.
\eeq
The expression (\ref{OmegaR}) allows for the right identification of the
angle conjugated to the $R$--symmetry \cite{BeKr}, 
\beq
\psi' = 3\tau + \phi + \psi ~.
\label{Rangle}
\eeq
Finally, starting from $J_4$, we can write the K\"ahler form $J$ of $\CLabc$,
\beq
J = r^2\,J_4 + r\,dr \wedge e^5\,\,, \qquad dJ = 0 ~.
\label{Kform}
\eeq
Notice that all the expressions written in this section reduce to those of
$\CYpq$ provided
\bear
&&a\,=\,p\,-\,q ~, \qquad b\,=\,p\,+\,q ~, \qquad c\,=\,p ~, \rc
&&3x\,-\,\alpha\,=\,2\,\alpha\,y ~, \qquad \mu\,=\,{4\over 27}\,(1\,-\,a)\,\alpha^3 ~, \\
&&\tilde\theta\,=\,2\theta ~, \qquad \tilde\beta\,=\,-\,(\phi\,+\,\psi) ~,
\qquad \tilde\phi\,=\,\phi\,-\,\psi ~, \nonumber
\eear
while (\ref{Rangle}) provides the right identification with the $U(1)_R$
angle in $\Ypq$.
We shall use this limiting case several times along the letter to make
contact with the results found in \cite{CaEdPaZaRaVa}.

\setcounter{equation}{0}
\section{Killing spinors for $AdS_5\times \Labc$}
\label{Kspinors}
\medskip

In order to study D--brane probes' embeddings by means of kappa symmetry, we
need to know the Killing spinors of the string theory background. The solution
of type IIB supergravity corresponding to the near-horizon region of a stack of $N$ coincident D3-branes located at the apex of the $\mathcal{C}\Labc$ cone,
is characterized by a ten-dimensional metric,
\beq
ds^2\,=\,{r^2\over L^2}\,dx^2_{1,3}\,+\,{L^2\over r^2}\,dr^2\,+\,
L^2\,ds^2_{\Labc}\,\,,
\label{10dmetric}
\eeq
and a self-dual Ramond-Ramond five-form $F^{(5)}$ given by:
\beq
g_s\,F^{(5)}=d^4x\,\wedge dh^{-1}\,+\,{\rm Hodge~dual}\,\,,
\qquad\quad h(r)={L^4\over r^4}\,\,.
\label{F5}
\eeq
The quantization condition of the flux of $F^{(5)}$ determines the constant
$L$ in terms of $g_s$, $N$, $\alpha'$ and the volume of the Sasaki--Einstein space:
\beq
L^4\,=\,{4\pi^4\over {\rm Vol}(\Labc)}\,g_s\,N\,(\alpha')^2\,\,.
\label{L}
\eeq
The Killing spinors of the $AdS_5\times \Labc$ background can be written
in terms of a constant spinor $\eta$,
\beq
\epsilon\,=\,e^{{i\over 2}(3\tau+\phi+\psi)}\,r^{-{\Gamma_{*}\over 2}}\,\,
\Big(\,1\,+\,{\Gamma_r\over 2L^2}\,\,x^{\alpha}\,\Gamma_{x^{\alpha}}\,\,
(1\,+\,\Gamma_{*}\,)\,\Big)\,\,\eta\,\,,
\label{adsspinor}
\eeq
where we have introduced the matrix $\Gamma_{*}\,\equiv
i\Gamma_{x^0x^1x^2x^3}$. The spinor $\eta$ satisfies the projections
\cite{SfZo}:
\beq
\Gamma_{12}\,\eta\,=\,i\eta\,\,, \qquad\qquad
\Gamma_{34}\,\eta\,=\,i\eta\,\,,
\label{etaspinor}
\eeq
this implying that $\epsilon$ also satisfies the same projections. It is
convenient to decompose the constant spinor $\eta$ according to its
$\Gamma_{*}$--parity, ~$\Gamma_{*}\,\eta_{\pm}\,=\,\pm\eta_{\pm}$.
Using this decomposition, we obtain two types of Killing spinors:
\bear
e^{-{i\over 2} (3\tau+\phi+\psi)}\,\epsilon_{-} & = & r^{1/2}\,\eta_-\,\,,\rc\rc
e^{-{i\over 2} (3\tau+\phi+\psi)}\,\epsilon_{+} & = & r^{-1/2}\,\eta_+\,+\,{r^{1/2}
\over L^2}\,\,\Gamma_r\,x^{\alpha}\, \Gamma_{x^{\alpha}}\,\eta_+\,\,.
\label{chiraladsspinor}
\eear
The spinors $\epsilon_-$ satisfy
$\Gamma_{*}\,\epsilon_-\,=\,-\epsilon_-$, whereas the $\epsilon_+$'s are
not  eigenvectors of $\Gamma_{*}$. The former correspond to  ordinary
supercharges while the latter,  which depend on the $x^\alpha$
coordinates, are related to  the superconformal supersymmetries. The only
dependence on the coordinates of $\Labc$ is through the exponential of
$\psi' = 3\tau +
\phi + \psi$. This angle, as explained above, is identified with the
$U(1)_R$ of the superconformal quiver theory.

It is finally convenient to present the explicit expression for the Killing
spinors when $AdS_5$ is described by its global coordinates,
\beq
ds^2_{AdS_5}\,=\,L^2\,\Big[ -\cosh^2\varrho \,\,dt^2\,+\,d\varrho^2\,+\,
\sinh^2\varrho\,\,d\Omega_3^2 \Big]\,\,,
\label{globalADS}
\eeq
where $d\Omega_3^2$ is the round metric of a unit three-sphere. Let us
parameterize $d\Omega_3^2$ in terms of three angles $(\alpha^1,
\alpha^2,\alpha^3)$ as 
$d\Omega_3^2\,=\,(d\alpha^1)^2\,+\,\sin^2\alpha^1\Big(\,
(d\alpha^2)^2\,+\,\sin^2\alpha^2\,(d\alpha^3)^2\,\Big)$. 
The Killing
spinors in these coordinates take the form:
\beq
\epsilon\,=\,e^{{i\over 2}(3\tau+\phi+\psi)}\,
e^{-i\,{\varrho\over 2}\,\Gamma_{\varrho}\gamma_*}\,
e^{-i\,{t\over 2}\,\Gamma_{t}\gamma_*}\,
e^{-{\alpha^1\over 2}\,\Gamma_{\alpha^1 \rho}}\,
e^{-{\alpha^2\over 2}\,\Gamma_{\alpha^2 \alpha^1}}\,
e^{-{\alpha^3\over 2}\,\Gamma_{\alpha^3 \alpha^2}}\,\eta\,\,,
\label{globalspinor}
\eeq
where $\gamma_*\equiv\,\Gamma_{t}\,\Gamma_{\varrho}\,
\Gamma_{\alpha^1\,\alpha^2\,\alpha^3}$ and 
$\eta$ is a constant spinor that satisfies the same conditions as 
in (\ref{etaspinor}). 

\setcounter{equation}{0}
\section{Quiver theories for $\Labc$ spaces}
\label{Quiver}
\medskip

The $\Labc$ superconformal field theories were first constructed in
\cite{BeKr,FrHaMaSpVeWe,BuFoZa}. They are four dimensional quiver theories
whose main features we would like to briefly remind. 
\begin{table}[ht]
\begin{center}
$$\begin{array}{|c|c|c|c|c|c|}
\hline
\mathrm{Field}&R -\mathrm{charge}&\mathrm{number}&
U(1)_B &  U(1)_{F_1} & U(1)_{F_2} \\
\hline\hline
& & & & & \\[-1ex]
Y  & {2\over 3}\,{x_3\,-\,x_1\over x_3} & b &  a &  1 & 0 \\[1.5ex]
\hline
& & & & & \\[-1ex]
Z  & {2\over 3}\,{x_3\,-\,x_2\over x_3} & a &  b &  0 & k \\[1.5ex]
\hline 
& & & & & \\[-1ex]
U_1 & {2\over 3}\,{\alpha\over x_3}     & d & -c &  0 & l \\[1.5ex]
\hline
& & & & & \\[-1ex]
U_2 & {2\over 3}\,{\beta\over x_3}      & c & -d & -1 & -k-l \\[1.5ex]
\hline
& & & & & \\[-1ex]
V_1 & {2\over 3}\,{2x_3\,+\,x_1\,-\,\beta\over x_3} & c-a & b-c & 0 &
k+l \\[1.5ex]
\hline 
& & & & & \\[-1ex]
V_2 & {2\over 3}\,{2x_3\,+\,x_1\,-\,\alpha\over x_3} & b-c & c-a & -1 &
-l \\[1.5ex]
\hline 
\end{array}$$
\caption{Charges for bifundamental chiral fields in the quiver dual to
$\Labc$ \cite{FrHaMaSpVeWe}.}
\label{charges}
\end{center}
\end{table}
The gauge theory for
$\Labc$ has $N_g = a + b$ gauge groups and $N_f = a + 3 b$ bifundamental
fields. The latter are summarized in Table~\ref{charges}. There is a
$U(1)_F^2$ flavor symmetry that corresponds, in the gravity side, to the
subgroup of isometries that leave invariant the Killing spinors. There is
a certain ambiguity in the choice of flavor symmetries in the gauge theory
side, as long as they can mix with the $U(1)_B$ baryonic symmetry group.
This fact is reflected in the appearance of two integers $k$ and $l$ in
the $U(1)_F^2$ charge assignments, whose only restriction is given by the
identity $c\,k\,+\,b\,l\,=\,1$ (here, it is assumed that $b$ and $c$ are
coprime) \cite{FrHaMaSpVeWe}.

The charge assignments in Table~\ref{charges} fulfill a number of
nontrivial constraints. For example, all linear anomalies vanish,
$\mathrm{Tr}\,U(1)_B =
\mathrm{Tr}\,U(1)_{F_1} = \mathrm{Tr}\,U(1)_{F_2} = 0$. The cubic t'~Hooft anomaly, $\mathrm{Tr}U(1)_B^3$, vanishes as well. The superpotential of the theory has three kind of terms; a quartic one,
\beq
Tr\, Y\,U_1\,Z\,U_2 ~,
\eeq
and two cubic terms,
\beq
Tr\, Y\,U_1\,V_2 ~, \qquad Tr\, Y\,U_2\,V_1 ~.
\eeq
Their R-charge equals two and they are neutral with respect to the baryonic
and flavor symmetries. The number of terms of each sort is uniquely fixed
by the multiplicities of the fields to be, respectively, $2\,a$, $2\,(b-c)$
and $2\,(c-a)$ \cite{FrHaMaSpVeWe}. The total number of terms, then, equals
$N_f - N_g$. In the $\Ypq$ limit, the isometry of the space --thus
the global flavor symmetry-- enhances, $U^1$ and $U^2$ (also $V^1$ and $V^2$)
becoming a doublet under the enhanced $SU(2)$ group. The superpotential
reduces in this limit to the $\Ypq$ expression \cite{BeFrHaMaSp}. More
details about the $\Labc$ superconformal gauge theories can be found in
\cite{BeKr,FrHaMaSpVeWe,BuFoZa}.

\setcounter{equation}{0}
\section{D3-branes on three-cycles}
\label{D3s}
\medskip

In this section we consider D3--brane probes wrapping three-cycles of $\Labc$.
These are pointlike objects from the gauge theory point of view, corresponding
to dibaryons constructed from the different bifundamental fields of the quiver
theory. There are other configurations of physical interest that we will not
discuss in this letter. Though, we will briefly discuss their most salient
features in Section 9.

Given a D3--brane probe wrapping a supersymmetric three-cycle $\CC$, the conformal dimension $\Delta$ of the corresponding dual operator is
proportional to the volume of the wrapped three-cycle,
\beq
\Delta\,=\,{\pi\over 2}\,{N\over L^3}\,\,
{{\rm Vol}({\cal C})\over {\rm Vol}({ \Labc})}\,\,.
\eeq
Since the $R$-charge of a protected operator is related to its dimension
by $R={2\over 3}\Delta$, we can readily compute the $R$-charge of the dibaryon
operators. On the other hand, the baryon number associated to the D3--brane probe wrapping $\CC$ (in units of $N$) can be obtained as the integral over
the cycle of the pullback of a $(2,1)$-form $\Omega_{2,1}$:
\beq
{\cal B}({\cal C})\,=\,\pm i \int_{{\cal C}}\,
P\big[\,\Omega_{2,1}\,\big]_{{\cal C}}\,\,.
\label{baryon-def}
\eeq
The explicit form of $\Omega_{2,1}$ is:
\beq
\Omega_{2,1}\,=\,{K\over  \rho^4}\,\Big(\,{dr\over r}\,+\,i\,
e^5\,\Big) \wedge \left(\,e^1\wedge e^2\,-\,e^3\,\wedge e^4\,\right)\,\,,
\label{Odosuno}
\eeq
where $K$ is a constant that will be determine below. Armed with these
expression, we can extract the relevant gauge theory information of the
configurations under study.

\subsection{$U_1$ dibaryons}
\medskip
\label{U1}

Let us take the worldvolume coordinates for the D3-brane probe to be
$\xi^{\mu}\,=\,(t,x,\psi,\tau)$, with $\theta\,=\,\theta_0$ and
$\phi\,=\,\phi_0$ constant, and let us assume that the brane is located
at a fixed point in $AdS_5$. The action of the kappa symmetry matrix on
the Killing spinor reads
\beq
\Gamma_{\kappa}\,\epsilon\,=\,-{i \over 4!\sqrt{ -\det g}}\,
\epsilon^{\mu_1\cdots\mu_4}\,\gamma_{\mu_1\cdots\mu_4}\,\epsilon\,=\,
-{iL^4 \over \sqrt{ -\det g}}\,
\big[\,a_5\Gamma_{t5}\,+\,a_{135}\,\Gamma_{t135}\,\big]\,\epsilon\,\,,
\label{U1-antisymmetrized}
\eeq
where
\beq
a_5\,=\,-i{\cosh\varrho\over 2\beta}\,\cos^2\theta\,\,,\qquad
a_{135}\,=\,-{\cosh\varrho\over 4 \sqrt{\Delta_x}}\,\,
\bigg(1-{x\over \beta}\bigg)\,\,\sqrt{\Delta_\theta}\,\sin (2\theta)\,\,.
\eeq
Compatibility of (\ref{U1-antisymmetrized}) with the projections
(\ref{etaspinor}) demands $a_{135}\,=\,0$.
Since $\Delta_\theta$ cannot vanish for positive $\alpha$ and $\beta$, this
condition implies $\sin(2\theta)\,=\,0$, \ie\, $\theta=0$ or $\pi/2$. Due to the
fact that, for these configurations, the determinant of the induced metric is:
\beq
-\det g\,=\,{L^8\over 4}\,\Bigg[\,
{\Delta_\theta\sin^2(2\theta)\over 4\Delta_x}\,
\bigg(1-{x\over \beta}\bigg)^2\,+\,{\cos^4\theta\over \beta^2}\,\Bigg]\,
\cosh^2\varrho\,\,,
\eeq
we must discard the $\theta=\pi/2$ solution since the volume of the cycle
would vanish in that case. Thus, the D3--brane probe is placed at
$\theta=0$ and the kappa symmetry condition
$\Gamma_{\kappa}\,\epsilon\,=\,\epsilon$ reduces to the new projection:
\beq
\Gamma_{t 5}\,\epsilon\,=-\,\epsilon\,\,,
\eeq
which can only be imposed at the center of $AdS_5$. The corresponding
configuration preserves four supersymmetries.

Given that the volume of ${\cal U}_1$ can be
easily computed with the result
\beq
{\rm Vol}(\,{\cal U}_1\,)\,=\,{\pi L^3\over \beta}\,
(x_2-x_1)\,{\Delta \tau\over k}\,\,,
\eeq
the corresponding value for the $R$-charge is:
\beq
R_{{\cal U}_1}\,=\,{2\over 3}\,{\alpha\over \alpha+\beta-x_1-x_2}\,N\,=\,
{2\alpha\over 3x_3}\,N\,\,,
\eeq
where we have used the second relation in (\ref{xs}). This result agrees with
the value expected for the operator $\det (U_1)$ \cite{FrHaMaSpVeWe}.
Let us now compute the baryon number associated to the D3--brane probe
wrapping ${\cal U}_1$. For the ${\cal U}_1$ cycle, we get
\beq
{\cal B}({\cal U}_1)\,=\,i\int_{{\cal U}_1}\,
P\big[\,\Omega_{2,1}\,\big]_{{\cal U}_1}\,=\,
-\,{2\pi^2\over \alpha\beta}\,{c\over a\,b}\,K\,\,,
\eeq
where we have used the second identity in (\ref{ratios2}). 
From the field theory analysis \cite{FrHaMaSpVeWe} it is known that the baryon
number of the $U_1$ field should be $-c$ (see Table \ref{charges}). We
can use this result to fix the constant $K$ to:
\beq
K\,=\,-\,{\alpha\beta\over 2\pi^2}\,a\,b\,\,.
\label{Kconstant}
\eeq
Once it is fixed, formulas (\ref{baryon-def}) and (\ref{Odosuno}) allow us to
compute the baryon number of any D3--brane probe wrapping a three-cycle.

\subsection{$U_2$ dibaryons}
\medskip

Let us again locate the D3-brane probe at a fixed point in $AdS_5$ and take the
following set of worldvolume coordinates $\xi^{\mu}\,=\,(t,x,\phi,\tau)$,
with constant $\theta\,=\,\theta_0$ and $\psi\,=\,\psi_0$. The kappa symmetry matrix now acts on the Killing spinor as
\beq
\Gamma_{\kappa} \epsilon\,=\,-{iL^4 \over \sqrt{ -\det g}}\,
\big[\,b_5\Gamma_{t5}\,+\,b_{135}\,\Gamma_{t135}\,\big]\,\epsilon\,\,,
\eeq
where
\beq
b_5\,=\,-i{\cosh\varrho\over 2\alpha}\,\sin^2\theta\,\,,\qquad
b_{135}\,=\,{\cosh\varrho\over 4 \sqrt{\Delta_x}}\,\,
\bigg(1-{x\over \alpha}\bigg)\,\,\sqrt{\Delta_\theta}\,\sin (2\theta)\,\,.
\eeq
The BPS condition is $b_{135}\,=\,0$, which can only be satisfied if $\sin (2\theta)=0$. We have to select now the solution $\theta\,=\,{\pi\over 2}$
if we want to have a non-zero volume for the cycle. The above condition
defines the ${\cal U}_2$ cycle. The associated R-charge can be computed as
above and reads:
\beq
R_{{\cal U}_2}\,=\,{2\beta\over 3x_3}\,N\,\,,
\eeq
in precise agreement with the gauge theory result \cite{FrHaMaSpVeWe}.
The baryon number reads
\beq
{\cal B}({\cal U}_2)\,=\,
i\int_{{\cal U}_2}\,P\big[\,\Omega_{2,1}\,\big]_{{\cal U}_2}\,=\,
-c\,{\beta\over \alpha}\,\,{(\alpha-x_1)(\alpha-x_2)\over
(\beta-x_1)(\beta-x_2)}\,\,,
\eeq
where we have used (\ref{Kconstant}) and, after the third identity in
(\ref{ratios2}), we get:
\beq
{\cal B}({\cal U}_2)\,=\,-d\,=\,-(a+b-c)\,\,,
\eeq
in agreement with the field theory result \cite{MaSp2} (see Table\,\ref{charges}). If we consider the
case $a = p-q$, $b = p+q$ and $c = p$, which amounts to $\Ypq$, 
a $U(1)$ factor
of the isometry group enhances to $SU(2)$ and these dibaryons are
constructed out of a doublet of bifundamental fields.

\subsection{$Y, Z$ dibaryons}
\medskip

We now take the following set of worldvolume coordinates $\xi^{\mu}\,=\,(t,\theta,\psi,\tau)$ and 
the embedding $x = x_0$ and $\psi'\,=\,\psi'_0$, where $\psi'_0$ is 
a constant and $\psi' = 3\tau + \phi + \psi$ is the angle conjugated to
the $U(1)_R$ charge. We implement this embedding in our coordinates by
setting
$\phi(\psi,\tau) = \psi'_0 - 3\tau - \psi$. In this case
\beq
\Gamma_{\kappa} \epsilon\,=\,-{iL^4 \over \sqrt{ -\det g}}\,
\big[\,c_3\Gamma_{t3}\,+\,c_5\Gamma_{t5}\,+\,c_{135}\,\Gamma_{t135}\,\big]\,
\epsilon\,\,,
\eeq
where
\bear
&&c_3\,=\,3i\,{\rho \cosh\varrho\over 2\alpha\beta}\,\,\sin(2\theta)\,
\sqrt{\Delta_x}\,\,,\rc\rc
&&c_5\,=\,i\,{\cosh\varrho\over 2\alpha\beta}\,\sin(2\theta)\,
\left( 3 x^2 - 2 (\alpha + \beta) x + \alpha \beta \right)\,,\rc\rc
&&c_{135}\,=\,{\cosh\varrho\over \alpha\beta}\,{\alpha\,\cos^2\theta\,
(1-3\sin^2\theta)\,-\beta\,\sin^2\theta\,(1-3\cos^2\theta)\over
\sqrt{\Delta_\theta}}\,\sqrt{\Delta_x}\,\,.
\eear
The BPS conditions are, as before, $c_3\,=\,c_{135}\,=\,0$. Clearly these conditions are satisfied only if $\Delta_x\,=\,0$, or, in other words, when
\beq
x\,=\,x_1\,,\,x_2\,\,.
\eeq
Notice that the value of $\psi'_0$ is undetermined. The induced volume
takes the form:
\beq
\sqrt{-\det g}\,\big|_{x=x_i}\,=\,{L^4\over 2\alpha\beta}\,
\left| 3 x_i^2 - 2 (\alpha + \beta) x_i + \alpha \beta \right|\,\sin(2\theta)\,
\cosh\varrho\,\,.
\eeq
As before, the compatibility with the $AdS_5$ SUSY requires that $\rho=0$. 
Let us denote by ${\cal X}_i$ the cycle with $x=x_i$. We get that the volumes are given by:
\beq
{\rm Vol}\big({\cal X}_i\big)\,=\,{\pi\over k\,\alpha\,\beta}\,
\left| 3 x_i^2 - 2 (\alpha + \beta) x_i + \alpha \beta \right|\,
\Delta\tau\, L^3\,\,.
\eeq
From this result we get the corresponding values of the $R$-charges,
namely:
\beq
R_{{\cal Y}}\,=\,{2N\over 3}\,{x_3-x_1\over x_3}\,\,,\qquad\qquad
R_{{\cal Z}}\,=\,{2N\over 3}\,{x_3-x_2\over x_3}\,\,,
\eeq
where ${\cal Y} = {\cal X}_1$ and ${\cal Z} = {\cal X}_2$.
Let us now compute the baryon number of these cycles. 
The pullback of the three-form $\Omega_{2,1}$ to the cycles with $x=x_i$ and
$\psi'=\psi'_0$ is:
\beq
P\big[\,\Omega_{2,1}\,\big]_{x=x_i}\,=\,
iK\,{\left( 3 x_i^2 - 2 (\alpha + \beta) x_i + \alpha \beta \right)\over
2\alpha\beta}\,\,
{\sin(2\theta)\over \rho^4}\,\,
d\theta\wedge d\psi\wedge d\tau\,\,,
\eeq
where $K$ is the constant written in (\ref{Kconstant}). We
obtain: 
\beq
{\cal B}({\cal X}_i)\,=\,-i\,\int_{{\cal X}_i}\,
P\big[\,\Omega_{2,1}\,\big]_{{\cal X}_i}\,=\,{\pi\over k\,\alpha\beta}\,K\,
{3 x_i^2 - 2 (\alpha + \beta) x_i + \alpha \beta\over
(\alpha\,-\,x_i)\,(\beta\,-\,x_i)}\,\Delta\tau\,\,.
\eeq
Taking into account the third identity in (\ref{ratios2}), we get:
\beq
{\cal B}({\cal Y})\,=\,a\,\,,\qquad\qquad
{\cal B}({\cal Z})\,=\,b\,\,,
\eeq
as it should \cite{FrHaMaSpVeWe} (see Table \ref{charges}). 

\subsection{Generalized embeddings}
\medskip

In this subsection we show that there are generalized embeddings of D3--brane
probes that can be written in terms of the local complex coordinates (\ref{zs})
as  holomorphic embeddings or divisors of $\CLabc$. Let us consider, for
example, $(t,x,\psi,\tau)$ as worldvolume coordinates and the ansatz
\beq
\theta=\theta(x,\psi)\,\,,\qquad\qquad
\phi=\phi(x,\psi)\,\,.
\label{D3-general-ansatz}
\eeq
This ansatz is a natural generalization of the one used in section \ref{U1}.
The case where the worldvolume coordinate $\psi$ is changed by $\phi$, can
be easily addressed by changing $\alpha \to \beta$ and $\theta \to \pi/2
- \theta$.
Putting the D3-brane at the center of $AdS_5$, we get that the kappa symmetry
condition is given by an expression as in (\ref{U1-antisymmetrized})
\beq
\Gamma_{\kappa} \epsilon\,=\,-{iL^4 \over \sqrt{ -\det g}}\,
\big[\,a_5\Gamma_{t5}\,+\,a_{135}\,\Gamma_{t135}\,\big]\,\epsilon\,\,,
\label{U1g-antisymmetrized}
\eeq
where $a_5$ and $a_{135}$ are now given by:
\bear
&&a_5\,=\,-{i\over 2}\,\bigg[\,{\cos^2\theta\over \beta}\,+\,
{\sin^2\theta\over \alpha}\,\phi_\psi\,+\,\sin(2\theta)\,\bigg\{
\bigg(\,1-\,{x\over \beta}\,\bigg)\,\theta_x\,-\,\bigg(\,1-\,{x\over
\alpha}\,\bigg)\, \big(\theta_x\phi_\psi-\theta_\psi\phi_x\big)\,\bigg\}
\,\bigg]\,\,,\rc\rc
&&a_{135}\,=\,-\sqrt{{\Delta_\theta\over \Delta_x}}\,
{\sin(2\theta)\over 4}\,\bigg[\,
1-\,{x\over \beta}-\bigg(\,1-\,{x\over \alpha}\,\bigg)\,\phi_\psi\,\bigg]+
\sqrt{{\Delta_x\over \Delta_\theta}}\,\bigg[\,
{\cos^2\theta\over \beta}\theta_x\,\,+\rc\rc
&&\qquad\qquad+\,{\sin^2\theta\over \alpha}\,
\big(\theta_x\phi_\psi-\theta_\psi\phi_x\big)\,\bigg]\,+\,
{i\over 2}\,\bigg[\,\sqrt{\Delta_x\Delta_\theta}\,\,{\sin(2\theta)\over
\alpha\beta}\,\phi_x\,-\,{\rho^2\over \sqrt{\Delta_x\Delta_\theta}}\,
\theta_{\psi}\,\bigg]\,\,.\qquad\qquad
\eear
The BPS condition $a_{135}=0$ reduces to the following pair of differential
equations:
\bear
&&{\cos^2\theta\over \beta}\theta_x+ {\sin^2\theta\over \alpha}\,
\big(\theta_x\phi_\psi-\theta_\psi\phi_x\big)\,=\,
{\Delta_\theta\over \Delta_x}\,
\bigg[\,
1-\,{x\over \beta}-\bigg(\,1-\,{x\over \alpha}\,\bigg)\,\phi_\psi\,\bigg]
\,\,{\sin(2\theta)\over 4}\,\,,\rc\rc
&&\rho^2\theta_{\psi}\,=\,
{\Delta_x\Delta_\theta\over \alpha\beta}\,\,\sin(2\theta)\, \phi_x\,\,.
\label{D3-BPS-general}
\eear
The integral of the above equations can be simply written as:
\beq
z_2\,=\,f(z_1)\,\,,
\eeq
where $z_1$ and $z_2$ are the local complex coordinates of $\CLabc$ and 
$f(z_1)$ is an arbitrary holomorphic function. Actually, if $\xi^{\mu}$ is an
arbitrary worldvolume coordinate, one has:
\beq
\partial_{\xi^{\mu}}\,z_2\,=\,f'(z_1)\,\partial_{\xi^{\mu}}\,z_1\,\,.
\eeq
One can eliminate the function $f$ in the above equation by considering the
derivatives with respect to two worldvolume coordinates $\xi^{\mu}$ and
$\xi^{\nu}$. One gets:
\beq
\partial_{\xi^{\mu}}\,\log z_2\,\,
\partial_{\xi^{\nu}}\,\log z_1\,=\,
\partial_{\xi^{\nu}}\,\log z_2\,\,
\partial_{\xi^{\mu}}\,\log z_1\,\,.
\label{infalible}
\eeq
Taking $\xi^{\mu}=x$ and $\xi^{\nu}=\psi$ in the previous equation and
considering that the other coordinates $\theta$ and $\phi$ entering $z_1$
and $z_2$ depend on $(x,\psi)$ (as in the ansatz (\ref{D3-general-ansatz})),
one can prove that (\ref{infalible}) is equivalent to the system of BPS equations (\ref{D3-BPS-general}).

We have checked that the Hamiltonian density of a static D3--brane probe of
the kind discussed in this Section satisfies a bound that is saturated when
the BPS equations (\ref{D3-BPS-general}) hold. This comes from the fact that,
from the point of view of the probes, these configurations can be regarded
as BPS worldvolume solitons. We have also checked that these generalized
embeddings are calibrated
\beq
P\Big[\,{1\over 2}\,J\wedge J\,\Big]_{{\cal D}}\,=\,
{\rm Vol}({\cal D})\,\,,
\label{calibration}
\eeq
where  ${\rm Vol}({\cal D})$ is the volume form of the divisor 
${\cal D}$, namely 
${\rm Vol}({\cal D})\,=\,r^3\,dr\wedge {\rm Vol}({\cal C})$. It is
important to remind at this point that supersymmetry holds locally but it
is not always true that a general embedding makes sense globally. We have
seen examples of this feature in $\Ypq$ \cite{CaEdPaZaRaVa}.

\setcounter{equation}{0}
\section{D5-branes}
\label{D5s}
\medskip

Let us consider a D5-brane probe that creates a codimension one defect on the
field theory. It represents a domain wall in the gauge theory side such that, when one crosses one of these objects, the gauge groups change and one passes from an ${\cal N}=1$ superconformal field theory to a cascading theory with fractional branes. The setup for the supergravity dual of this cascading
theory was proposed in \cite{MaSp}.

We choose the following set of worldvolume coordinates: $\xi^{\mu}\,=\,(t,x^1,x^2,r, \theta,\phi)$, and we will adopt the ansatz $x=x(\theta,\phi)$,
$\psi\,=\,\psi(\theta,\phi)$, $\tau\,=\,\tau(\theta,\phi)$
with $x^3$ constant. The kappa
symmetry matrix acts on the spinor $\epsilon$ as:
\beq
\Gamma_\kappa\,\epsilon\,=\,{i\over \sqrt{-\det g}}\,{r^2\over L^2}\,
\Gamma_{x^0x^1x^2r}\,\gamma_{\theta\phi}\,\epsilon^*\,=\,
{i\over \sqrt{-\det g}}\,r^2\,\Gamma_{x^0x^1x^2r}\,\big[\,
b_I\,+\,b_{15}\,\Gamma_{15}\,+\,b_{35}\,\Gamma_{35}\,+\,b_{13}
\,\Gamma_{13}\,\big]\,
\epsilon^*\,\,,
\label{D5gamathetaphi}
\eeq
where 
\bear
&&b_I\,=\,{i\over 2}\,\bigg[\,\sin(2\theta)\,\bigg(\,1-\,{x\over \alpha}\,-
\bigg(\,1-\,{x\over \beta}\,\bigg)\,\psi_{\phi}\,\bigg)\,-\,
{\sin^2\theta\over \alpha}\,x_{\theta}\,+\,{\cos^2\theta\over \beta}\,
\big(\,\psi_\theta\, x_\phi\,-\,\psi_\phi\,
x_\theta\,\big)\,\bigg]\,\,,\rc\rc\rc
&&b_{15}\,=\,{\rho\over \sqrt{\Delta_{\theta}}}\,
\bigg[\,\bigg(\,1-\,{x\over \alpha}\,\bigg)\,\sin^2\theta+
\bigg(\,1-\,{x\over \beta}\,\bigg)\cos^2\theta\,\psi_{\phi}\,
+\,\tau_{\phi}\,\bigg]
\,\,-\rc\rc
&&\qquad
-{i\over 2}\,\sin(2\theta)\,{\sqrt{\Delta_{\theta}}\over \rho}
\,\,\Bigg[\,\,
\bigg(\,1-\,{x\over \alpha}\,\bigg)
\bigg[\,\tau_{\theta}\,+\,\bigg(\,1-\,{x\over
\beta}\,\bigg)\,\psi_\theta\,\bigg]\,+\,
\bigg(\,1-\,{x\over \beta}\,\bigg)\bigg(\,
\tau_{\phi}\,\psi_{\theta}\,-\,\tau_{\theta}\,\psi_{\phi}\,\bigg)
\,\,\Bigg]
\,,\rc\rc\rc
&&b_{35}\,=\,{\sqrt{\Delta_x}\over \rho}\,\Bigg[\,\,
{\alpha-\beta\over 4\alpha\beta}\,
\sin^2(2\theta)\,\psi_{\theta}\,-\,{\sin^2\theta\over
\alpha}\,\tau_{\theta}\,+\,{\cos^2\theta\over \beta}\,
\bigg(\,\tau_{\phi}\,\psi_{\theta}\,-\,\tau_{\theta}\,\psi_{\phi}\,\bigg)
\,\,\Bigg]
\,\,+\rc\rc
&&\qquad
+\,{i\over 2}\,{\rho\over \sqrt{\Delta_x}}\,
\bigg[\bigg(\,1-\,{x\over \alpha}\,\bigg)\,\sin^2\theta\,x_\theta\,-\,
\bigg(\,1-\,{x\over \beta}\,\bigg)\cos^2\theta
\big(\,\psi_\theta\, x_\phi\,-\,\psi_\phi\, x_\theta\,\big)\,
+\,\tau_{\phi}\,x_{\theta}\,-\,\tau_{\theta}\,x_{\phi}
\,\bigg]\,\,,\rc\rc
&&b_{13}\,=\,{1\over 4}\,\sqrt{{\Delta_\theta\over \Delta_x}}\,
\sin(2\theta)\,\bigg[\,
\bigg(\,1-\,{x\over \alpha}\,\bigg)\,x_\theta\,+\,
\bigg(\,1-\,{x\over \beta}\,\bigg)
\big(\,\psi_\theta\, x_\phi\,-\,\psi_\phi\, x_\theta\,\big)\,\bigg]\,\,+
\rc\rc
&&\qquad
+\,\sqrt{{\Delta_x\over \Delta_\theta}}\,\bigg[\,
{\sin^2\theta\over \alpha}\,+\,{\cos^2\theta\over \beta}\,\psi_{\phi}\,
\bigg]\,+\,{i\over 2}\,\bigg[\,{\rho^2\over  \sqrt{\Delta_x\Delta_\theta}}\,
x_{\phi}\,-\,\sqrt{\Delta_x\Delta_\theta}\,\,
{\sin(2\theta)\over \alpha\beta}\,\psi_{\theta}\,\bigg]\,\,.
\label{D5-coefficients}
\eear
The BPS conditions are $b_I\,=\,b_{15}\,=\,b_{35}\,=\,0$. The imaginary part
of $b_{15}$ is zero if $\psi_{\theta}=\tau_{\theta}=0$, \ie,
$\psi\,=\,\psi(\phi)$, $\tau\,=\,\tau(\phi)$. Let us assume that this is
the case and define the quantities $n$ and $m$ as:
\beq
\psi_{\phi}\,=\,n\,\,,
\qquad\qquad
\tau_{\phi}\,=\,m\,\,.
\label{D5-psi-tau}
\eeq
Clearly $n$ and $m$ are independent of the angle $\theta$. Moreover, from
the vanishing of the real part of $b_{15}$ and of the imaginary part of 
$b_{35}$ we get an algebraic equation for $x$, which can be solved as:
\beq
x\,=\,\alpha\beta\,\,{\sin^2\theta\,+\,n\cos^2\theta\,+\,m
\over\beta\sin^2\theta \,+\,n\alpha\cos^2\theta}\,\,.
\label{D5-xtheta}
\eeq
On the other hand, when $\psi_\theta=0$  and $\psi_{\phi}\,=\,n$ the
vanishing of $b_I$ is equivalent to the equation:
\beq
\bigg[\,{\sin^2\theta\over \alpha}\,+\,{\cos^2\theta\over \beta}\,
n\,\bigg]\,x_{\theta}\,=\,
\sin(2\theta)\,\bigg(\,1-\,{x\over \alpha}\,-
\bigg(\,1-\,{x\over \beta}\,\bigg)\,n\,\bigg)\,\,,
\eeq
which is certainly satisfied by our function 
(\ref{D5-xtheta}). For an embedding satisfying (\ref{D5-psi-tau}) and
(\ref{D5-xtheta}) one can check that 
$\sqrt{-\det g}\,=\,r^2\,|\,b_{13}\,|$. Therefore, for these embeddings, 
$\Gamma_{\kappa}$ acts on the Killing spinors as:
\beq
\Gamma_{\kappa}\,\epsilon\,=\,i e^{i\delta_{13}}\,\Gamma_{x^0x^1x^2r}\,
\Gamma_{13}\,\epsilon^*\,\,,
\eeq
where $\delta_{13}$ is the phase of $b_{13}$. In order to implement
correctly  the kappa symmetry condition
$\Gamma_{\kappa}\,\epsilon\,=\,\epsilon$, the phase $\delta_{13}$ must be
constant along the worldvolume of the probe. By inspecting the form of
the coefficient $b_{13}$  in (\ref{D5-coefficients}), one readily
concludes that $b_{13}$ must be real, which happens only when
$x_{\phi}=0$. Moreover, it follows from (\ref{D5-xtheta}) that $x$ is
independent of $\phi$ only when $n$ and $m$ are constant. Thus, $\psi$
and $\tau$ are linear functions of the angle $\phi$, namely:
\beq
\psi\,=\,n\phi\,+\,\psi_0\,\,,
\qquad\qquad
\tau\,=\,m\phi\,+\,\tau_0\,\,,
\eeq
where $\psi_0$ and $\tau_0$ are constant. Notice that in these
conditions the equation 
$\Gamma_{\kappa}\,\epsilon\,=\,\epsilon$ reduces to 
\beq
i \,\Gamma_{x^0x^1x^2r}\,
\Gamma_{13}\,\epsilon^*\,=\,\epsilon\,\,.
\label{D5-condition}
\eeq
Due to the presence of the complex conjugation, (\ref{D5-condition})
is only consistent if the R-charge angle $\psi'=3\tau+\phi+\psi$ is
constant along the worldvolume (see the expression of $\epsilon$ in
(\ref{adsspinor})). This in turn gives rise to an additional
restriction to the possible supersymmetric embeddings. Indeed, the
condition 
$3\tau+\phi+\psi\,=\,\psi'_0\,=\,{\rm constant}$ implies that the
constants $n$ and $m$ satisfy
\beq
3m+n+1\,=\,0\,\,.
\eeq
Thus, the possible supersymmetric  embeddings of the D5-brane are
labeled by a constant $n$ and are given by:
\bear
&&\psi\,=\,n\phi\,+\,\psi_0\,\,,
\qquad\qquad
\tau\,=\,-{n+1\over 3}\,\,\phi\,+\,\tau_0\,\,,\rc\rc
&&x\,=\,{\alpha\beta\over 3}\,\,
{2-n-3(1-n)\,\cos^2\theta\over \beta\,+\,
(n\alpha\,-\,\beta)\,\cos^2\theta}\,\,.
\eear
It can be now checked as in refs. \cite{ArCrRa,CaEdPaZaRaVa} that the
projection (\ref{D5-condition}) can be converted into a set of  algebraic
conditions on the constant spinors $\eta_{\pm}$ of
(\ref{chiraladsspinor}). These conditions involve a projector which
depends on the constant R-charge angle $\psi_0'=3\tau_0+\psi_0$ and has
four possible solutions. Therefore these embeddings are 1/8
supersymmetric.

The configuration obtained in this section can be also shown to saturate
a Bogomol'nyi bound in the worldvolume theory of the D5--brane probes. This
amounts to a point of view in which the solution is seen as a worldvolume
soliton. 

Other configurations of physical interest can be considered at this point.
Most notably, we expect to find stable non-supersymmetric configurations of
D5--branes wrapping three cycles of $\Labc$. A similar solution where the
D5--brane probe wraps the entire $\Labc$ manifold, thus corresponding to the
baryon vertex of the gauge theory, should also be found. We will not include
the detailed analysis of these aspects in this article.

\setcounter{equation}{0}
\section{Spacetime filling  D7-branes}
\label{D7s}
\medskip

Let us consider a D7--brane probe that fills the four Minkowski gauge theory
directions while possibly extending along the holographic direction. These
configurations are relevant to add flavor to the gauge theory. In particular,
the study of fluctuations around them provides the meson spectrum. We start
from the following set of worldvolume coordinates $\xi^{\mu}\,=\, (x^0,x^1,x^2,x^3,x,\psi,\theta,\phi)$ and the ansatz $r\,=\,r(x,\theta)$,
$\tau=\tau(\psi,\phi)$. The kappa symmetry matrix in this case reduces to:
\beq
\Gamma_{\kappa}\,\epsilon\,=\,-i\,{r^4\over L^4\sqrt{-\det g}}\,
\Gamma_{x^0\cdots x^3}\,\gamma_{x\psi\theta\phi}\,\epsilon\,\,.
\eeq
Let us assume that the Killing spinor $\epsilon$ satisfies the condition 
$\Gamma_*\epsilon=-\epsilon$, \ie\ $\epsilon$ is of the form $\epsilon_-$ and,
therefore, one has:
\beq
\Gamma_{x^0\cdots x^3}\,\epsilon_-\,=\,i\epsilon_-\,\,,
\label{D3chirality}
\eeq
which implies $\Gamma_{r5}\epsilon_-\,=\,i\epsilon_-$. Then:
\beq
\Gamma_{\kappa}\,\epsilon_-\,=\,{r^4\over \sqrt{-\det g}}\,
\big[\,d_I\,+\,d_{15}\,\Gamma_{15}\,+\,d_{35}\,\Gamma_{35}\,+\,
d_{13}\Gamma_{13}\,\big]\,\epsilon_-\,\,.
\label{D7gamma-epsilon-}
\eeq
In order to express these coefficients in a compact form, let us define
$\Lambda_x$ and $\Lambda_{\theta}$ as:
\bear
&&\Lambda_x\,=\,-{1\over 2\Delta_x}\bigg[\,\big(\,\alpha-x
\,\big)\,
\big(\,\beta-x\,\big)\,+\,\alpha\big(\,\beta-x\,\big)
\tau_{\phi}\,+\,\beta\big(\,\alpha-x\,\big)\tau_{\psi}\,\bigg]\,
\,,\rc\rc
&&\Lambda_\theta\,=\,{1\over \Delta_\theta}\bigg[\,
\big(\alpha-\beta\big)\,\,
\sin\theta\cos\theta\,+\,\alpha\,\cot\theta\,\,\tau_{\phi}\,-\,
\beta\tan\theta\,\,\tau_{\psi}\,\bigg]\,\,.
\label{d7Lambda}
\eear
Then:
\bear
&&d_I\,=\,{\sin\theta\cos\theta\over 2\alpha\beta}\,\,
\bigg[\,\rho^2\,+\,\Delta_\theta\,\,\Lambda_\theta\,{r_\theta\over r}\,+\,
4\,\Delta_x\,\,\Lambda_x\,{r_x\over r}\,\bigg]\,\,,\rc\rc
&&d_{15}\,=\,i\rho\,\,
{\sin\theta\cos\theta\over 2\alpha\beta}\,\,\sqrt{\Delta_{\theta}}\,\,\bigg[\,
{r_\theta\over r}\,-\,\Lambda_{\theta}\,\bigg]\,\,,\rc\rc
&&d_{35}\,=\,-\rho \,\,{\sin\theta\cos\theta\over \alpha\beta}\,\,
\sqrt{\Delta_{x}}\,\,\,\,\bigg[\,
{r_x\over r}\,-\,\Lambda_{x}\,\bigg]\,\,,\rc\rc
&&d_{13}\,=\,i\,{\sin\theta\cos\theta\over \alpha\beta}\,\,
\sqrt{\Delta_{\theta}\Delta_{x}}\,\,\,\bigg[\,
\Lambda_{x}\,{r_\theta\over r}\,-\,\Lambda_{\theta}\,
{r_x\over r}\,\bigg]\,\,. 
\eear
The BPS conditions are clearly $d_{15}\,=\,d_{35}\,=\,d_{13}\,=\,0$.
From the vanishing of $d_{15}$ and $d_{35}$ we get the following
first-order equations:
\beq
{r_\theta\over r}\,=\,\Lambda_\theta\,\,,\qquad\qquad
{r_x\over r}\,=\,\Lambda_x\,\,.
\label{d7BPS}
\eeq
Notice that $d_{13}=0$ as a consequence of these equations. 
By looking at the explicit form of our ansatz and at the
expression of $\Lambda_\theta$ and $\Lambda_x$ in (\ref{d7Lambda}), one realizes
that the only dependence on the angles $\phi$ and $\psi$ in the first-order
equations (\ref{d7BPS}) comes from the partial derivatives of $\tau(\psi,\phi)$.
For consistency these derivatives must be constant, \ie\
$\tau_{\psi}\,=\,n_{\psi}$, $\tau_{\phi}\,=\,n_{\phi}$, where $n_{\psi}$
and $n_{\phi}$ are constants. These equations can be trivially integrated:
\beq
\tau(\psi,\phi)\,=\,n_{\psi}\,\psi\,+\,n_{\phi}\,\phi\,+\,\tau_0\,\,.
\label{d7tau}
\eeq
Notice that $\tau(\psi,\phi)$ relates angles whose periods are not congruent.
Thus, the D7--brane spans a submanifold that is not, in general, a cycle.
It is worth reminding that this is not a problem for flavor branes. 
If the BPS conditions (\ref{d7BPS}) hold one can check that 
$r^4d_I\,=\,\sqrt{-\det g}$ and, therefore, one has indeed that 
$\Gamma_{\kappa}\epsilon=\epsilon$ for any Killing spinor
$\epsilon=\epsilon_-$, with $\epsilon_-$ as in (\ref{chiraladsspinor}).
Thus these configurations preserve the four ordinary supersymmetries
of the background.

In order to get the dependence of $r$ on $\theta$ and $x$ it is
interesting to notice that, if $\tau(\psi, \phi)$ is given by
(\ref{d7tau}), the integrals of 
$\Lambda_{\theta}$ and $\Lambda_{x}$ turn out to be:
\bear
&&\int\,\Lambda_{\theta}\,d\theta\,=\,\log
\Bigg[\,{(\sin\theta)^{n_\phi}\,(\cos\theta)^{n_\psi}\over
\Delta_{\theta}^{{n_\phi+n_\psi+1\over 2}}}\,\Bigg]\,\,,\rc\rc
&&\int\,\Lambda_{x}\,dx\,=\,\log
\Bigg[\,{[f_1(x)]^{{n_\phi-n_\psi\over 2}}\over
\Delta_x^{{1\over 6}}\,\,
[f_2(x)]^{{n_\phi+n_\psi\over 2}+{1\over 3}}}\,\Bigg]\,\,,
\eear
where $f_1(x)$ and $f_2(x)$ are the functions defined in (\ref{f12}). 
From this result it straightforward to obtain the general solution of 
$r(\theta,x)$:
\beq
r(\theta,x)\,=\,C\,
{(\sin\theta)^{n_\phi}\,(\cos\theta)^{n_\psi}\over
\Delta_{\theta}^{{n_\phi+n_\psi+1\over 2}}}\,\,\,
{[f_1(x)]^{{n_\phi-n_\psi\over 2}}\over
\Delta_x^{{1\over 6}}\,\,
[f_2(x)]^{{n_\phi+n_\psi\over 2}+{1\over 3}}}\,\,,
\label{r-theta-x}
\eeq
where $C$ is a constant. Notice that the function $r(x,\theta)$ diverges for some particular values of $\theta$ and $x$. This means that the probe always extends infinitely in the holographic direction. For particular values of
$n_\phi$ and $n_\psi$ there is a minimal value of the coordinate $r$,
$r_\star$, which depends on the integration constant $C$. If one uses
this probe as a flavor brane, $r_\star$ provides an energy scale that
is naturally identified with the mass of the dynamical quarks added to the
gauge theory.

It is finally interesting to write the embedding characterized by eqs. 
(\ref{d7tau}) and (\ref{r-theta-x}) in terms of the complex coordinates $z_1$,
$z_2$ and  $z_3$ defined in eq. (\ref{zs}). Indeed, one can check that this
embedding can be simply written as:
\beq
z_1^{m_1}\,z_2^{m_2}\,z_3^{m_3}\,\,=\,{\rm constant}\,\,,
\eeq
where $m_3\not=0$. The relation between the exponents $m_i$ and the 
constants $n_\psi$ and $n_\phi$ is the following:
\beq
{m_1\over m_3}\,=\, {3\over 2}\,\,(\,n_\psi\,-\,n_\phi)\,\,,
\qquad\qquad
{m_2\over m_3}\,=\, -{3\over 2}\,\,(\,n_\psi\,+\,n_\phi)\,-\,1\,\,.
\eeq
By using  the Dirac--Born--Infeld
action of the D7--brane, it is again possible to show that  there
exists a bound for the energy which is saturated for BPS configurations.

\setcounter{equation}{0}
\section{Final Remarks}
\label{FinalRemarks}
\medskip

In this letter we have worked out supersymmetric configurations involving D--brane probes in $AdS_5 \times \Labc$. Our study focused on three kinds of branes, namely D3, D5 and D7. We have dealt with embeddings corresponding to dibaryons, defects and flavor branes in the gauge theory. For
D3--branes wrapping three-cycles in $\Labc$ we first reproduced all quantum
numbers of the bifundamental chiral fields in the dual quiver theory. We
also found a new class of supersymmetric embeddings of D3--branes in this background that we identified with a generic holomorphic embedding. The
three-cycles wrapped by these D3--branes are calibrated. In the case of
D5--branes, we found an embedding that corresponds to a codimension one
defect.
From the point of view of the D5--branes, it can be seen as a BPS saturated
worldvolume soliton. We finally found a spacetime filling D7--brane probe
configuration that can be seen to be holomorphically embedded in the
Calabi--Yau, and is a suitable candidate to introduce flavor in the
quiver theory.

Other interesting configurations have been considered following the lines
of \cite{CaEdPaZaRaVa}. We would only list their main features:

{\it Fat strings} ~If we take a D3--brane with worldvolume coordinates
$(x^0,x^1,\theta,\phi)$ and consider an embedding of the form $x =
x(\theta,\phi)$ and $\psi = \psi(\theta,\phi)$, with the remaining scalars
constant, we see that there is no solution preserving kappa symmetry.
However, we have obtained a fat string solution by wrapping a probe
D3--brane on a two-cycle, which is the same considered in Section 7
for a D5--brane probe. This configuration is not supersymmetric but it
is stable.

{\it D5 on a three-cycle} ~We have found an embedding corresponding to
D5--branes that wrap a three-cycle in $\Labc$. They are codimension one
in the gauge theory coordinates. These configurations happen to be non
supersymmetric yet stable.

{\it D5 on a two-cycle} ~We studied another embedding where a D5--brane
probe wraps a two-cycle in $\Labc$ while it extends along the radial
coordinate. For this embedding, $\phi$, $\psi$, $x^3$ and $\tau$ are
held constant. This is a supersymmetric configuration. We also considered  
turning on a worldvolume flux  in the case studied in Section \ref{D5s},
and found that it can be done in a supersymmetric way. The flux in the 
worldvolume of the brane provides a bending  of the profile $x^3$ of the
wall, analogously to what happens in $T^{1,1}$ \cite{ArCrRa} and $\Ypq$
\cite{CaEdPaZaRaVa}.

{\it Another spacetime filling D7} ~We considered a different spacetime
filling D7--brane that extends infinitely in the radial direction and wraps
a three-cycle holomorphically embedded in $\Labc$ of the type studied
in Section 6.4. It preserves four supersymmetries.

{\it D7 on $\Labc$} ~We finally studied a D7--brane probe wrapping the
entire $\Labc$ space and extended along the radial coordinate. From the
point of view of the gauge theory, this is a string-like configuration
that preserves two supersymmetries.

It would be interesting to study in more detail the introduction of flavor
in these theories and, in particular, to compute the corresponding meson
spectra. These results exhaust the study of D--brane probes at the tip of
toric Calabi--Yau cones on a base with topology $S^2 \times S^3$
initiated in
\cite{ArCrRa,CaEdPaZaRaVa}.

\medskip
\section*{Acknowledgments}
\medskip
We are pleased to thank useful comments from Sebasti\'an Franco and Dario
Martelli.
This work was supported in part by MCyT, FEDER and Xunta de Galicia under
grant FPA2005-00188 and by  the EC Commission under grants HPRN-CT-2002-00325
and MRTN-CT-2004-005104. JDE is also supported by the FCT grant
POCTI/FNU/38004/2001.  
Institutional support to the Centro de Estudios Cient\'\i ficos (CECS) from
Empresas CMPC is gratefully acknowledged. CECS is a Millennium Science
Institute and is funded in part by grants from Fundaci\'on Andes and the
Tinker Foundation.



\end{document}